\documentclass[aps,prd,notitlepage,nofootinbib,superscriptaddress, twocolumn]{revtex4}
\usepackage{amssymb}
\usepackage{color}
\usepackage{amsmath}
\usepackage{verbatim}
\usepackage{appendix}
\usepackage[usenames,dvipsnames]{xcolor}

\newcommand{\fs}{{\cal F}_s}
\newcommand{\gs}{{\cal G}_s}

\begin{document}

\title{Generic analysis of kinetically driven inflation}

\author{Rio Saitou}
\email{rsaito@ntu.edu.tw}
\affiliation{Leung Center for Cosmology and Particle Astrophysics, National Taiwan University, Taipei, Taiwan 10617}
\affiliation{School of Physics, Huazhong University of Science and Technology, Wuhan 430074, China}

\begin{abstract}

We perform a model-independent analysis of kinetically driven inflation (KDI) which (partially) includes Generalized G-inflation and Ghost inflation. We evaluate the background evolution splitting into the inflationary attractor and the perturbation around it. We also consider the quantum fluctuation of scalar mode with a usual scaling and derive the spectral index ignoring the contribution from the second order products of slow-roll parameters. Using these formalisms, we find that within our generic framework, the models of KDI which possess the shift symmetry of scalar field cannot create the quantum fluctuation consistent to the observation. Breaking the shift symmetry, we obtain a few essential conditions for viable models of KDI associated with the graceful exit.
\end{abstract}

\maketitle

\section{Introduction}\label{sec1}

Inflation is an elegant scenario which can resolve the problems in the big bang universe by accelerating the early universe \cite{Guth:1980zm, Starobinsky:1980te, Sato:1980yn}. 
The quantum fluctuations produced in the inflationary universe become the seed of large-scale structure, and their nearly scale-invariant, Gaussian features \cite{Mukhanov:1981xt, Starobinsky:1982ee, Hawking:1982cz, Guth:1982ec} are compatible with the precise observation of cosmic microwave background anisotropies \cite{Planck:2013jfk, Ade:2015lrj}.
Now, it is not too much say that inflation is the standard model of the early universe, while we still have some alternatives \cite{Steinhardt:2001st, Mukohyama:2009gg, Creminelli:2010ba}. 

In the inflationary models, slow-roll inflation, in which a scalar field rolls down on its potential slowly, is known as a common class of the models \cite{Albrecht:1982wi, Linde:1983gd, Steinhardt:1984jj, Martin:2013tda}. In that class, the potential energy of scalar field plays a role of the vacuum energy to accelerate the universe. Typically, slow-roll inflation ends naturally and changes to the big bang universe after the scalar field reaches to a smaller value than the Planck mass. 
Taking a few stringent constraints from the observation of cosmic microwave background, the form of scalar potential has been narrowed down until now.

To begin with, however, are there any reasons why inflation is driven by the scalar potential? In reality, we do have another huge class of accelerating the universe, so-called kinetically driven inflation (KDI) \cite{ArmendarizPicon:1999rj, Garriga:1999vw, ArkaniHamed:2003uz, Kobayashi:2010cm, Burrage:2010cu, Kobayashi:2011nu}. In this class, 
the kinetic terms of scalar field play an important role to inflate the universe.
KDI can also transit to the big bang universe by tuning functions of the scalar field in the models. 
After the first model of KDI \cite{ArmendarizPicon:1999rj} was proposed, many models of this class have been united in the frame of Horndeski theory \cite{Kobayashi:2011nu, Horndeski:1974wa}. Although the quantum fluctuation behaves differently from the Horndeski theory, ghost inflation \cite{ArkaniHamed:2003uz} holds the same mechanism for inflating the universe as KDI. 
Some models of KDI exhibit features in sharp contrast with slow-roll inflation \cite{Garriga:1999vw, ArkaniHamed:2003uz, Kobayashi:2010cm, Burrage:2010cu, Babich:2004gb, Chen:2006nt, Kobayashi:2011pc, Gao:2011qe, DeFelice:2011uc, DeFelice:2013ar}, and thus it is expected that the precise observations in the future can distinguish which class has much concordance with the observational data.

So far, however, any unified analytical methods for describing the whole evolution of KDI have not been developed. 
The main purpose of this article is to reformulate the previous courses of analyses and develop a unified formulation for evaluating the behavior of KDI systematically. 
We perform a generic analysis of background evolution splitting into the inflationary attractor and linear perturbations around it explicitly. We derive the equation of motion for the linear perturbations with an ansatz where all derivatives with respect to the scalar field $\phi$ are sufficiently small to be treated as perturbations. 
Further, we consider the scalar quantum fluctuation around the background with a usual scaling dimension. Using a well-known approximation where we keep the first order slow-roll parameters only,
we classify the asymptotic form of the mode function by the rank of Hankel function. 
Using those formalisms, we analyze the behavior of KDI for the shift symmetric case and $\phi$-dependent case separately. Then, we find a few model-independent constraints with which viable models of KDI should be satisfied. 

This article is organized as follows. In the next section, we provide a theoretical framework in a model-independent manner and show the inflationary attractor of KDI with an explicit example. In Sec. \ref{3}, we analyze the background evolution and the quantum fluctuation perturbatively. We apply those formalisms to the shift symmetric case of KDI in Sec. \ref{4} and to $\phi$-dependent case in Sec. \ref{5}. Sec. \ref{6} is devoted for the conclusion.  
We add Appendix A and B, where technical details are provided. In Appendix \ref{C}, we shortly review a non-attractor solution, so-called ultra slow-roll solution \cite{Kinney:2005vj, Hirano:2016gmv}, which the theory can hold potentially.

\section{Attractor solution of kinetically driven inflation}\label{2}

In this article, we intend to derive model-independent features of kinetically driven inflation.
Thus, instead of specifying a model, we just impose the following conditions to a theory;  
\begin{itemize}
  \item The theory preserves a symmetry under 4-dimensional diffeomorphism.
  \item The action of theory consists of a single scalar field $\phi$ and a metric field $g_{\mu\nu}$, i.e., $S[\phi,\, g_{\mu\nu}]$.
  \item The theory contains one scalar mode and two tensor modes only in the focused energy scale.
  \item The theory has the flat Friedmann-Robertson-Walker(FRW) background solution.
\end{itemize}
We give the flat FRW metric as  
$ds^2 = -N(t)^2dt^2 + a(t)^2d\vec{x}^2$
where $N(t)$ is the lapse function. By variation of the action with respect to $N(t)$ and the background field $\phi=\phi(t)$, we can always get the following form of constraint equation for the background
\begin{equation}
\label{Feq}
    \mathcal{E}(\phi,\,\dot{\phi},\, H,\, \cdots)=0 \ ,
\end{equation} 
and the equation of motion for $\phi(t)$
\begin{align}
    \label{phi}
    \dot{J}+3HJ=P_{,\phi}  \ , 
\end{align}
where
\begin{align}
\label{j}
    J&=J(\phi,\,\dot{\phi},\, H,\, \cdots), \nonumber \\
    P &= P(\phi,\,\dot{\phi},\, \ddot{\phi},\, H,\, \dot{H},\,\cdots) \ .
\end{align}
The dot denotes the derivative with respect to time $t$ and $ H\equiv\dot{a}/a$ is the Hubble parameter. The ellipses imply contributions from the higher order time derivatives of $\phi$ and $H$, which are assumed to be negligible throughout this article. The functions ${\cal E},\ J$ and $P$ are supposed to be regular for their arguments. For simplicity, we restrict ${\cal E}$ and $J$ as they include up to the first order time derivatives of $\phi$ and $a$ as the valid terms. 
In this article, a comma followed with subscript(s) denotes a partial derivative with respect to the subscript(s), e.g., $P_{,\phi}= \partial P/\partial\phi$.
The constraint equation (\ref{Feq}) is equivalent to the Friedmann equation. 
For the derivation of these equations, see Appendix \ref{A}.

If the derivative with respect to $\phi$ in the right hand side of Eq. (\ref{phi}) are sufficiently small to exclude it from a leading evolution, or, the theory has a shift symmetry under $\phi\rightarrow \phi +c$ where $c$ is a constant parameter,
the equation of motion for $\phi$ is reduced to 
\begin{equation}
\label{att}
\dot{J} + 3HJ \simeq 0\ .
\end{equation}
A general solution for $J$ is 
\begin{equation}
\label{J}
J \simeq c_Ja^{-3}\ ,
\end{equation}
where $c_J$ is a constant of integration. When 
the universe expands, $J$ approaches to $0$ in the asymptotic future. 
If the field velocity $\dot{\phi}$ approaches to a nonzero value as $J\rightarrow 0$,
we could obtain a non-trivial quasi-de Sitter attractor solution.
Assume that the simultaneous equation system ${\cal E}=0$ and $J=0$ has at least one root solution as
\begin{equation}
\label{atts}
\dot{\phi}_0 = f(\phi_0)\neq 0,\quad H_0 = H_0(\phi_0) \ ,
\end{equation} 
where $f$ is a function of $\phi$ and the subscript $0$ stands for the leading order approximation. This root solution can be regarded as an attractor since the motion of background scalar field is determined by the first order differential equation.
If the Hubble parameter varies sufficiently slowly with respect to $\phi_0$, the  root solution \eqref{atts} can become a quasi-de Sitter attractor. Inflation induced by this mechanism is referred to as kinetically driven inflation.
As an explicit example, we consider the following model \cite{ArmendarizPicon:1999rj}
\begin{equation}
\label{exa}
S = \int d^4x \sqrt{-g}\left[ \frac{M_{\rm Pl}^2}{2}R - K(\phi)X + X^2\right] ,
\end{equation}
where $M_{\rm Pl}$ is the Planck mass, $R$ is the scalar curvature and $X:= -\frac{1}{2}g^{\mu\nu}\partial_\mu\phi\partial_\nu\phi$. For this action, we get the constraint equation
\begin{equation}
\label{ }
{\cal E} = -KX+3X^2 - 3M_{\rm Pl}^2H^2=0\ ,
\end{equation}
and the equation of motion for $\phi$
\begin{align}
\label{ }
&\dot{J} + 3HJ = -K_{,\phi} X, \nonumber \\
&J = \dot{\phi}(-K+2X) \ .
\end{align}
When $K>0$ and its derivative term with respect to $\phi$ is enough small to ignore, we obtain an attractor solution
\begin{equation}
\label{ }
\dot{\phi}_0 = \pm \sqrt{K(\phi_0)},\quad H_0= \frac{K(\phi_0)}{\sqrt{3}M_{\rm Pl}}\ .
\end{equation}
In the region where the function $K$ varies sufficiently slowly, the solution becomes a quasi-de Sitter attractor.

As we mentioned, the attractor (\ref{atts}) is the leading order approximation of full solution, and the actual motion of scalar field deviates from the attractor. In the following sections, we evaluate the deviation perturbatively.

\section{Perturbations around the KDI attractor}\label{3}

To derive the post-attractor solution, we consider isotropic and homogeneous perturbations around the attractor solution. We derive equations of motion for the perturbations using the full equations (\ref{Feq}) and (\ref{phi}). Further, we consider the quantum fluctuations around the background solution. 

\subsection{Perturbative solution of background}

We expand the background scalar field and the Hubble parameter as
\begin{equation}
\label{pert}
\phi(t) = \phi_0(t) + \phi_1(t),\quad H(t)= H_0(t) + h_1(t) \ .
\end{equation}
$\phi_1$ and $h_1$ are the linear perturbations around the attractor solution.
Introducing a small parameter $|\xi|\ll1$, we make an ansatz that \textit{all} of derivatives with respect to $\phi$ are as small as the linear perturbations:
\begin{align}
\label{ansatz}
&\left\{\phi_1,\,h_1,\,A_{,\phi}\right\} = O(\xi), \\
&A= \sum^\infty_{n=0}\left(\frac{\partial^n}{\partial \phi^n}\right)A_n, 
\nonumber
\end{align}
where $\{A_n\}$ are arbitrary functions of the background fields without differentiation with respect to $\phi$.
Then, the constraint equation (\ref{Feq}) is expanded up to $O(\xi)$ as
\begin{align}
\label{Fex}
\mathcal{E}&= \mathcal{E}_0  + \left(\mathcal{E}_{,\dot{\phi}}\right)_0\dot{\phi}_1+ \left(\mathcal{E}_{,H}\right)_0h_1 = 0,\\
{\cal E}_0 &:= {\cal E}(\phi_0, \dot{\phi}_0, H_0, \cdots), \nonumber 
\end{align}
where we ignored contributions from the higher order time derivative terms of $\phi$ and $H$.
The bracket with subscript $0$ implies that we substitute the leading values of background fields into the derivative in the bracket. Since the leading part ${\cal E}_0$ vanishes by itself, we obtain a relation between the perturbations
\begin{equation}
\label{h1}
h_1= -\left(\frac{\mathcal{E}_{,\dot{\phi}}}{\mathcal{E}_{,H}}\right)_0\dot{\phi}_1 .
\end{equation}
Similarly, the quantity $J$ is expanded up to $O(\xi)$ as
\begin{align}
\label{}
    J&=J_0 + \left(J_{,\dot{\phi}}\right)_0\dot{\phi}_1 + \left(J_{,H}\right)_0h_1, \\
 J_0&:=J(\phi_0\,,\dot{\phi}_0\,,H_0\,,\cdots) . \nonumber
\end{align}
The leading term $J_0$ vanishes, so we get
\begin{equation}
\label{j1}
J= \left\{ \left(J_{,\dot{\phi}}\right)_0 - \left(J_{,H}\frac{\mathcal{E}_{,\dot{\phi}}}{\mathcal{E}_{,H}}\right)_0\right\}\dot{\phi}_1, 
\end{equation}
where we used Eq. (\ref{h1}).
Substituting Eq. (\ref{j1}) into Eq. (\ref{phi}), we obtain the equation of motion for $\phi_1$ as
\begin{align}
\label{phi1}
&\ \ \ddot{\phi}_1 + 3H_0\dot{\phi}_1 = \left(\widetilde{P_{,\phi}}\right)_0, \\
&\widetilde{P_{,\phi}}:=\frac{P_{,\phi}}{\left(J_{,\dot{\phi}}\right)_0 - \left(J_{,H}\frac{\mathcal{E}_{,\dot{\phi}}}{\mathcal{E}_{,H}}\right)_0} .\nonumber
\end{align}
This equation is similar to the equation of motion for a canonical scalar field in the FRW spacetime.
Note that to make use of Eq. (\ref{phi1}), we must impose a condition 
\begin{equation}
\label{small}
|\dot{\phi}_0|\gg|\dot{\phi}_1|\ ,
\end{equation}
otherwise we can not treat $\phi_1$ as the linear perturbation around the KDI attractor (\ref{atts}).

Before solving the equation of motion for $\phi_1$, we consider the quantum fluctuation of scalar mode around the background in the next subsection.

\subsection{Quantum fluctuation}

We assume that after solving all of the constraints the theory has, we obtain the second order action for the scalar fluctuation $\zeta$ as
\begin{align}
\label{s2}
    S_2^{(S)}&= \int dtd^3xa^3\left[ {\cal G}_s\dot{\zeta}^2 - \frac{{\cal F}_s}{a^2}(\vec{\nabla} \zeta)^2 + \cdots\right]  \ , \\
    \fs&={\cal F}_s(\phi,\,\dot{\phi},\, \ddot{\phi},\, H\,, \dot{H},\,\cdots), \nonumber \\
     \gs&= {\cal G}_s(\phi,\,\dot{\phi},\,\ddot{\phi},\, H,\,\dot{H},\,\cdots) \ .  
\end{align}
We may regard $\zeta$ as the comoving curvature perturbation. 
The ellipses in the functions $\fs$ and $\gs$ denote the negligible higher derivative terms of $\phi$ and $H$, while the ellipsis in the action (\ref{s2}) represents contributions from the higher derivative terms of $\zeta$. If the higher derivative terms of $\zeta$ become relevant, the scaling of quantum fluctuation would change significantly\footnote{Ghost inflation \cite{ArkaniHamed:2003uz} is included in this case. Thus, by using our formalisms, we can analyze the background evolution of ghost inflation, but cannot the evolution of quantum fluctuation of the model.}, which requires individual analysis for each model.
Thus, we do not take such cases into account.
The generic action (\ref{s2}) covers, at least, all classes of Horndeski theory and DHOST theory \cite{Gleyzes:2013ooa, Gleyzes:2014dya, Gleyzes:2014qga, Gleyzes:2014rba, Langlois:2017mxy}. For the derivation of Eq. (\ref{s2}), see Appendix. \ref{B} and \cite{Tanaka:2012wi} also.
Using the background equations (\ref{h1}) and (\ref{phi1}), we expand $\fs$ and $\gs$ up to $O(\xi)$ as 
\begin{align}
\label{fs}
    \fs&={\fs}_0 + \left({\fs}_{,\dot{\phi}}\right)_0\dot{\phi}_1 + \left({\fs}_{,\ddot{\phi}}\right)_0\ddot{\phi}_1 \nonumber \\
    &\quad + \left({\fs}_{,H}\right)_0h_1 + \left({\fs}_{,\dot{H}}\right)_0\dot{h_1} 
     \nonumber \\
    &= {\fs}_0 + 3H_0\left({\fs}_1-{\fs}_2\right)\dot{\phi}_1 + {\fs}_2\left(\widetilde{P_{,\phi}}\right)_0,   \\
    \label{gs}
    \gs&= {\gs}_0 + \left({\gs}_{,\dot{\phi}}\right)_0\dot{\phi}_1 + \left({\gs}_{,\ddot{\phi}}\right)_0\ddot{\phi}_1 \nonumber \\
    &\quad + \left({\gs}_{,H}\right)_0h_1 + \left({\gs}_{,\dot{H}}\right)_0\dot{h_1} 
     \nonumber \\
     &= {\gs}_0 + 3H_0\left({\gs}_1-{\gs}_2\right)\dot{\phi}_1 + {\gs}_2\left(\widetilde{P_{,\phi}}\right)_0,
\end{align}
where
\begin{align}
\label{ }
&{\fs}_0:= \fs(\phi_0,\,\dot{\phi}_0,\,H_0,\,\cdots), \nonumber \\
&{\fs}_1:= \frac{1}{3H_0}\left\{\left({\fs}_{,\dot{\phi}}\right)_0 - \left({\fs}_{,H} 
                \frac{\mathcal{E}_{,\dot{\phi}}}{\mathcal{E}_{,H}}\right)_0\right\},  \nonumber \\
&{\fs}_2:= \left({\fs}_{,\ddot{\phi}}\right)_0 - \left({\fs}_{,\dot{H}}
                \frac{\mathcal{E}_{,\dot{\phi}}}{\mathcal{E}_{,H}}\right)_0, \nonumber \\
 &{\gs}_0:=  \gs(\phi_0,\,\dot{\phi}_0,\,H_0,\,\cdots), \nonumber \\
 & {\gs}_1:= \frac{1}{3H_0}\left\{\left({\gs}_{,\dot{\phi}}\right)_0 
              - \left({\gs}_{,H}\frac{\mathcal{E}_{, \dot{\phi}}}{\mathcal{E}_{,H}}\right)_0
               \right\}, \nonumber \\
  &{\gs}_2:= \left({\gs}_{,\ddot{\phi}}\right)_0 - 
                \left({\gs}_{,\dot{H}}\frac{\mathcal{E}_{,\dot{\phi}}}
                {\mathcal{E}_{,H}}\right)_0.
\end{align}
The square of speed of sound for $\zeta$ is defined as
\begin{align}
\label{ }
c_s^2 &:= \frac{\fs}{\gs} \nonumber \\
&= \frac{{\fs}_0 + 3H_0\left({\fs}_1-{\fs}_2\right)\dot{\phi}_1 + {\fs}_2\left(\widetilde{P_{,\phi}}\right)_0}{{\gs}_0 + 3H_0\left({\gs}_1-{\gs}_2\right)\dot{\phi}_1 + {\gs}_2\left(\widetilde{P_{,\phi}}\right)_0}\  .
\end{align}
For the fluctuation being stable, $\fs$ and $\gs$ should be in the range
\begin{equation}
\label{stability}
\fs \geq 0, \quad \gs >0,\quad \gs \geq \fs,
\end{equation}
which leads to $0\leq c_s^2 \leq 1$.
Then, we define slow-roll parameters as
\begin{align}
\label{sr}
    \epsilon_1&:= -\frac{\dot{H}}{H^2},\quad \epsilon_2:= \frac{\dot{\epsilon}_1}{H\epsilon_1}, \nonumber \\
     f_{s1}&:= \frac{\dot{\fs}}{H\fs},\quad f_{s2}:= \frac{\dot{f}_{s1}}{Hf_{s1}}, \\
      g_{s1}&:= \frac{\dot{\gs}}{H\gs}, \quad g_{s2}:= \frac{\dot{g}_{s1}}{Hg_{s1}} ,\nonumber
\end{align}
and generalized conformal time as
\begin{equation}
\label{ }
d\tau_s := \frac{c_s}{a}dt .
\end{equation}
%
The derivative of $c_s/aH$ with respect to $\tau_s$ yeilds 
\begin{equation}
\label{csahd}
   \frac{d}{d\tau_s}\left(\frac{c_s}{aH}\right) = -1+\epsilon_1+\frac{1}{2}(f_{s1}-g_{s1}).  
\end{equation} 
If we regard the slow-roll parameters $\epsilon_1$, $f_{s1}$ and $g_{s1}$ as constants, 
the integral of Eq. (\ref{csahd}) reduces to
\begin{equation}
\label{csah}
\frac{c_s}{aH}\simeq -\left\{1- \epsilon_1-\frac{1}{2}(f_{s1}-g_{s1})\right\}\tau_s .
\end{equation}
This approximation implies practically that we ignore contributions from the second order products of slow-roll parameters $\epsilon_1\epsilon_2$, $f_{s1}f_{s2}$ and $g_{s1}g_{s2}$. In fact, from Eq. (\ref{csahd}), we obtain
\begin{align}
\label{tau}
   \tau_s &= -\int^{\tau_s} \frac{\frac{d}{d\tau_s'}\left(\frac{c_s}{aH}\right)}{\left\{1-
                    \epsilon_1-\frac{1}{2}(f_{s1}-g_{s1})\right\}}d\tau_s' \nonumber \\
              &= -\frac{c_s}{aH}\frac{1}{\left\{1- \epsilon_1-\frac{1}{2}(f_{s1}-g_{s1})\right\}} \nonumber \\
              &\quad + \int^{\tau_s}\frac{\epsilon_1\epsilon_2
                   + \frac{1}{2}(f_{s1}f_{s2}-g_{s1}g_{s2})}{\left\{1- \epsilon_1-\frac{1}{2}(f_{s1}-g_{s1})\right\}^2}d\tau_s' \ .
\end{align}
To get the relation (\ref{csah}), we need to discard the integral consisting of $\epsilon_1\epsilon_2$, $f_{s1}f_{s2}$ and $g_{s1}g_{s2}$ in the last line of Eq. (\ref{tau}). 
We always employ this approximation. In addition, we impose a condition
\begin{equation}
\label{taut}
1-\epsilon_1-\frac{1}{2}(f_{s1}-g_{s1}) > 0 
\end{equation}
which enforces $\tau_s$ to run from $-\infty$ to $0$ when the universe expands.

Discarding the contribution from $\epsilon_1\epsilon_2$, $f_{s1}f_{s2}$ and $g_{s1}g_{s2}$, we obtain the equation of motion for the scalar fluctuation in Fourier space as
\begin{align}
\label{zeta}
    &\partial_{\tau_s}^2\zeta_k - \frac{2\nu_s -1}{\tau_s} \partial_{\tau_s}\zeta_k+ k^2\zeta_k=0 ,    \\
    \label{nus}
    &\nu_s:= \frac{3-\epsilon_1+ g_{s1}}{2-2\epsilon_1-f_{s1}+g_{s1}} .
\end{align}
The linearly independent solutions of Eq. (\ref{zeta}) are 
\begin{align}
\label{}
    \zeta_k= &\frac{1}{2}\sqrt{\frac{\pi}{2}}(-k\tau_s)^{3/2}\frac{H\left(1-\epsilon_1-\frac{1}{2}(f_{s1}-g_{s1})\right)\gs^{1/4}}{k^{3/2}\fs^{3/4}} \nonumber \\
   &\times H_{\nu_s}^{(1)}(-k\tau_s)    
\end{align}
and its complex conjugate, where $H_{\nu_s}^{(1)}$ is the Hankel function of the first kind. We normalized the solution so that deep inside the horizon, the mode function $\zeta_k$ behaves as in Minkowski spacetime:
\begin{equation}
\label{ }
\zeta_k\approx \frac{1}{\sqrt{2}a(\fs\gs)^{1/4}}\frac{\mathrm{e}^{-ik\tau_s}}{\sqrt{2k}} \qquad \text{for}\quad |k\tau_s|\rightarrow \infty.
\end{equation}
On superhorizon scales where $|k\tau_s| \ll 1$, the asymtotic form of $\zeta_k$ for $\nu_s>0$ is 
\begin{align}
\label{ }
|\zeta_k| \simeq &\frac{1}{2}(-k\tau_s)^{3/2-\nu_s}2^{\nu_s-3/2}\frac{\Gamma(\nu_s)}{\Gamma(3/2)}\nonumber \\
&\times \frac{H\left(1-\epsilon_1-\frac{1}{2}(f_{s1}-g_{s1})\right)\gs^{1/4}}{k^{3/2}\fs^{3/4}} .
\end{align}
The scalar spectrum and the spectral index are obtained as 
\begin{align}
\label{ }
&P_\zeta := \frac{k^3}{2\pi^2}|\zeta_k|^2 = \frac{\gamma_s}{2}(-k\tau_s)^{3-2\nu_s}\frac{\gs^{1/2}}{\fs^{3/2}}\frac{H^2}{4\pi^2} , \\
&n_s-1:= \frac{d\text{ln}P_\zeta}{d\text{ln}k} = 3-2\nu_s,
\end{align}
where $\gamma_s:= 2^{2\nu_s-3}|\Gamma(\nu_s)/\Gamma(\frac{3}{2})|^2\left(1-\epsilon_1-\frac{1}{2}(f_{s1}-g_{s1})\right)^2$.
For $\nu_s<0$, the asymptotic form of $\zeta_k$ on superhorizon scales is
\begin{align}
\label{ }
        |\zeta_k| &\simeq \frac{|A|}{2}(-k\tau_s)^{3/2+\nu_s}\frac{H\left(1-\epsilon_1-\frac{1}{2}(f_{s1}-g_{s1})\right)\gs^{1/4}}{k^{3/2}\fs^{3/4}},  \\
        A &:= \sqrt{\frac{\pi}{2}}2^{-\nu_s}\left[ \frac{1}{\Gamma(1+\nu_s)} + \frac{{\rm cos}(\pi\nu_s)\Gamma(-\nu_s)}{\pi i}\right] \nonumber .
\end{align}
In this case, the scalar spectrum and the spectral index are obtained as
\begin{align}
\label{}
&P_\zeta = \frac{|A|^2\left(1-\epsilon_1-\frac{1}{2}(f_{s1}-g_{s1})\right)^2}{2}(-k\tau_s)^{3+2\nu_s}\nonumber \\
&\qquad\ \times \frac{\gs^{1/2}} {\fs^{3/2}}\frac{H^2}{4\pi^2} ,\\
\label{nm}
&n_s-1 = 3+2\nu_s .
\end{align}
For $\nu_s=0$, the asymptotic form of $\zeta_k$ on superhorizon scales becomes
\begin{align}
\label{}
    |\zeta_k|\simeq &\sqrt{\frac{1}{2\pi}}(-k\tau_s)^{3/2}{\rm ln}(-k\tau_s)\nonumber \\
    &\times \frac{H\left(1-\epsilon_1-\frac{1}{2}(f_{s1}-g_{s1})\right)\gs^{1/4}}{k^{3/2}\fs^{3/4}}  .
\end{align}
The scalar spectrum and the spectral index are obtained as
\begin{align}
\label{}
    P_\zeta =& \frac{\left(1-\epsilon_1-\frac{1}{2}(f_{s1}-g_{s1})\right)^2}{\pi}(-k\tau_s)^{3}\left\{{\rm ln}(-k\tau_s)\right\}^2\nonumber \\
    &\times \frac{\gs^{1/2}} {\fs^{3/2}}\frac{H^2}{4\pi^2},   \\
    n_s-1 &= 3 + \frac{2}{{\rm ln}(-k\tau_s)} \simeq 3  .
\end{align}
We note that if $\nu_s\leq0$ or, equivalently, $3-\epsilon + g_{s1}\leq0$, the scalar fluctuation grows even on superhorizon scales. 
The two independent solutions of $\zeta$ in real space on superhorizon scales are obtained as
\begin{align}
\label{}
    \zeta= \text{const.} \quad \text{and} \quad \int^t \frac{dt'}{\gs a^3}.
\end{align}
We denote the latter solution as the decaying mode.
The decaying mode can be expressed as
\begin{align}
\label{}
    \zeta_{\rm decay}\propto\begin{cases}
      & a^{-(3-\epsilon + g_{s1})}\quad \text{for}\quad 3-\epsilon + g_{s1}\neq0 \\
      & {\rm ln}\,a \ \,\qquad\qquad \text{for} \quad 3-\epsilon + g_{s1}=0\ .
\end{cases} 
\end{align}
Thus, the decaying mode does decay if $3-\epsilon_1+g_{s1}>0$, while it grows actually if $3-\epsilon_1+g_{s1}\leq0$. 
%

Using these formalisms, we evaluate the behavior of background and quantum fluctuation dividing into two cases in the following sections.
\section{Case I: shift symmetric system}\label{4}

First, we consider a simpler case that the action has a shift symmetry under $\phi \rightarrow \phi+c$, where $c$ is a constant parameter.
Provided at least one root solution of the KDI attractor, we get a $\phi$-\textit{independent} attractor solution
\begin{equation}
\label{}
    \dot{\phi}_0= c_*= \text{const}  \neq 0,\quad H_0 = \text{const} > 0,
\end{equation}
which reflects the shift symmetry of the action. 
Since $\phi$-dependence gets lost by the shift symmetry, Eq. (\ref{phi1}) reduces to 
\begin{equation}
\ddot{\phi}_1 + 3H_0\dot{\phi}_1= 0 .
\end{equation}
The solution is
\begin{equation}
\label{phi1a}
\dot{\phi}_1 \simeq \frac{c_1}{a^3}, 
\end{equation}
where $c_1$ is a constant of integration. The condition (\ref{small}) can be easily satisfied by choosing $c_1$ properly.
The behavior of perturbation $\phi_1$ is almost the same as the scalar field in the original ultra slow-roll inflation \cite{Kinney:2005vj} (See Appendix.). Therefore, when the theory possesses the shift symmetry, the KDI attractor becomes an exact de Sitter solution, and the linear perturbation around it behaves like ultra slow-roll inflation. 


Then, we consider features of the quantum fluctuation for the shift symmetric system. Using Eq. (\ref{phi1a}), we obtain the functions (\ref{fs}) and (\ref{gs}) as
\begin{align}
\label{}
    \fs& \simeq {\fs}_0 + 3H_0c_1\left({\fs}_1-{\fs}_2\right)a^{-3}  \ ,  \\
    \gs& \simeq {\gs}_0 +3H_0c_1\left({\gs}_1-{\gs}_2\right)a^{-3} \ ,
\end{align}
and the speed of sound as
\begin{align}
\label{}
    c_s^2 \simeq \frac{{\fs}_0 + 3H_0c_1\left({\fs}_1-{\fs}_2\right)a^{-3}}{{\gs}_0 + 3H_0c_1\left({\gs}_1-{\gs}_2\right)a^{-3}}   \ .
\end{align}
We note that for the shift symmetric system, all the functions but $a$ in $c_s^2$ become constants.
We can classify the shift symmetric system into three types depending on the values of ${\fs}_0$ and ${\gs}_0$: 
(i) ${\gs}_0\geq{\fs}_0 > 0$
  (ii) ${\fs}_0=0$ and ${\gs}_0 > 0$
 (iii) ${\fs}_0={\gs}_0=0$.
We do not consider any other types since they do not satisfy the stability  
condition (\ref{stability}).
For the type (i), after several e-folds, we eventually reach to a phase in which
\begin{align}
\label{ }
{\fs}_0 &\gg 3H_0c_1\left({\fs}_1-{\fs}_2\right)a^{-3}, \nonumber \\
{\gs}_0 &\gg 3H_0c_1\left({\gs}_1-{\gs}_2\right)a^{-3} \ .
\end{align}
Then, the slow-roll parameters (\ref{sr}) almost behave as
\begin{align}
\label{}
    \epsilon_1\propto f_{s1}\propto g_{s1} \propto a^{-3} , \quad \epsilon_2 \simeq f_{s2}\simeq g_{s2}\simeq -3 \ .  
\end{align}
We obtain $O(1)$ values for $\epsilon_2,\ f_{s2}$ and $g_{s2}$. 
In this case, the contribution from the second order products $\epsilon_1\epsilon_2$, $f_{s1}f_{s2}$ and $g_{s1}g_{s2}$ in Eq. (\ref{tau}), which are discarded in our formalism, becomes comparable with the contribution from $\epsilon_1$, $f_{s1}$ and $g_{s1}$:
\begin{align}
\label{lead}
&\int^{\tau_s}\frac{\epsilon_1\epsilon_2
                   + \frac{1}{2}(f_{s1}f_{s2}-g_{s1}g_{s2})}{\left\{1- \epsilon_1-\frac{1}{2}(f_{s1}-g_{s1})\right\}^2}d\tau_s' \nonumber \\
                  \simeq &\int^{\tau_s}\left\{-3\epsilon_1- \frac{3}{2}(f_{s1}-g_{s1})\right\}d\tau_s' \nonumber \\
                  \simeq &\frac{3c_s}{4aH}\left\{\epsilon_1 +\frac{1}{2}(f_{s1}-g_{s1})\right\}\ ,
\end{align}
where we picked up the leading part of integral only.
Therefore, we must dismiss $\epsilon_1$, $f_{s1}$ and $g_{s1}$ also from Eq. (\ref{tau}).  This results in
\begin{equation}
\label{ }
\quad\nu_s = \frac{3}{2},\quad n_s-1 = 0.
\end{equation}
 Thus, we have no scalar spectral tilt for the type (i) theories\footnote{In \cite{Hirano:2016gmv}, the spectral tilt of a Galileon model included into the type (i) theories is derived as proportional to $\epsilon_1$. However, we should ignore the tilt if we employ the approximation where we discard $\epsilon_1\epsilon_2\simeq -3\epsilon_1$.
}
as well as the original ultra slow-roll inflation \cite{Kinney:2005vj}. In the case of type (i) theories, however, the scalar fluctuation does never grow on superhorizon scales unlike ultra slow-roll inflation since $\nu_s$ for the type (i) theories is always positive, while $\nu_s$ for ultra slow-roll inflation would be negative (See Appendix \ref{C}). 
 
%

For the type (ii), after several e-folds, we get
\begin{equation}
\label{ }
{\fs}_0 = 0,\quad {\gs}_0 \gg 3H_0c_1\left({\gs}_1-{\gs}_2\right)a^{-3} \ .
\end{equation}
The slow-roll parameters almost behave as
\begin{align}
\label{}
   \epsilon_1 \propto g_{s1}\propto a^{-3} , \quad \epsilon_2 \simeq f_{s1} \simeq g_{s2} = -3, \quad f_{s2}= 0 . 
\end{align}
Similarly to the type (i), we must dismiss $\epsilon_1$ and $g_{s1}$ from Eq. (\ref{tau}) while keeping $f_{s1}$. This leads to the following spectral tilt
\begin{equation}
\label{ii}
\nu_s = \frac{3}{5},\quad n_s-1 = \frac{9}{5}.
\end{equation}
Thus, we have the blue spectrum for the type (ii) theories. 

Lastly, for the type (iii) where ${\fs}_0 = {\gs}_0 = 0$, we get the following slow-roll parameters
\begin{align}
\label{}
   \epsilon_1 \propto a^{-3} , \quad \epsilon_2 \simeq f_{s1} = g_{s1} = -3,\quad f_{s2}=g_{s2}= 0  .
\end{align}
We dismiss $\epsilon_1$ from Eq. (\ref{tau}) while keeping $f_{s1}$ and $g_{s1}$. The spectral tilt reduces to 
\begin{equation}
\label{ }
\nu_s = 0,\quad n_s-1 = 3 , 
\end{equation}
and we have the blue spectrum for the type (iii) theories also. 
Since $\nu_s=0$, the scalar fluctuation of the type (iii) theories grows logarithmically in terms of $a$ on superhorizon scales. 

Given these results, we find that the shift symmetric models of KDI within our generic framework cannot create the scalar fluctuation satisfied with $n_s\sim0.968$ if the higher order derivatives of $\zeta$ in the action remain irrelevant or do not exist. For example, the action of scalar fluctuation of Horndeski theory is described as Eq. (\ref{s2}) \cite{Kobayashi:2011nu}, so that all the shift symmetric models of KDI in Horndeski theory can not explain the observation. 
To obtain viable models of KDI, we need to introduce other sources for the scalar fluctuation, or, break the shift symmetry \cite{ArmendarizPicon:1999rj}. In the next section, we consider the case where the shift symmetry is broken, but the leading background evolution is still illustrated by the KDI attractor \eqref{atts}.

\section{Case II: $\phi$-dependent system}\label{5}

We break the shift symmetry and accept the system to depend on $\phi$. This case has been considered in the previous models to a limited extent \cite{ArmendarizPicon:1999rj, ArkaniHamed:2003uz, Kobayashi:2010cm, Burrage:2010cu}. In this section, we perform a model-independent analysis for this case. 

Provided at least one root solution of the KDI attractor depending on $\phi_0$, 
we obtain a leading solution (\ref{atts}) and the eqution of motion for $\phi_1$ (\ref{phi1}):
\begin{align}
\label{}
    &\dot{\phi}_0= f(\phi_0),\quad H_0=H_0(\phi_0), \nonumber  \\
     &\ddot{\phi}_1+3H_0\dot{\phi}_1=(\widetilde{P_{,\phi}})_0  \ . \nonumber
\end{align}
If $|\ddot{\phi_1}| \ll |(\widetilde{P_{,\phi}})_0|$, we obtain an approximate solution as
%
\begin{equation}
\label{ }
\dot{\phi}_1 \simeq  \frac{(\widetilde{P_{,\phi}})_0}{3H_0} \ .
\end{equation}
This indicates that the linear perturbation of $\phi$-dependent system ``slow-rolls" around the quasi-de Sitter KDI attractor. 
Thus, if we can neglect the second order time derivatives of $\phi_1$, the whole evolution of background is determined by $\phi$-dependence of $f$, $H_0$ and $(\widetilde{P_{,\phi}})_0$  similarly to slow-roll inflation.
The slow-roll condition for the perturbation follows from $|\ddot{\phi_1}| \ll |(\widetilde{P_{,\phi}})_0|$ up to $O(\xi)$ as
\begin{align}
\label{eta}
    \eta_\phi:= \frac{f(P_{,\phi\phi})_0}{3H_0(P_{,\phi})_0},\quad |\eta_\phi|\ll 1  \ .
\end{align}
By imposing $|\dot{\phi}_0|\gg|\dot{\phi}_1|$, 
we obtain another condition as 
\begin{align}
\label{ephi}
    \epsilon_\phi:= \frac{(P_{,\phi})_0}{3fH_0\left\{ (J_{,\dot{\phi}})_0 - \left(J_{,H}\frac{\mathcal{E}_{,\dot{\phi}}}{\mathcal{E}_{,H}}\right)_0\right\}},\quad |\epsilon_\phi| \ll 1 .   
\end{align}
The condition $|\epsilon_\phi|\ll 1$ generalizes the equivalent conditions in k-inflation and Ghost inflation. 
Further, to obtain a quasi-de Sitter expansion, we need the following condition
\begin{align}
\label{e1}
\epsilon_1 &\simeq -\frac{f}{H_0^2}\left\{ {H_0}_{,\phi_0} - 3H_0\left(\frac{\mathcal{E}_{,\dot{\phi}}}{\mathcal{E}_{,H}}\right)_0\epsilon_\phi\eta_\phi \right\}\nonumber \\
                   &\simeq -\frac{f{H_0}_{,\phi_0}}{H_0^2},\nonumber \\
|\epsilon_1|&\ll 1 \ ,
\end{align}
where we ignored $\epsilon_\phi\eta_\phi$ since it is the second order product of small parameters.
For the $\phi$-dependent system of KDI, the above three conditions are essential to realizing a quasi-de Sitter expansion mimicking slow-roll inflation. 

Due to the $\phi$-dependence of system, we could stop the acceleration dynamically if the slow-roll parameter $\epsilon_1$ varies to unity
\begin{equation}
\label{e11}
|\epsilon_1|\simeq \left|\frac{f{H_0}_{,\phi_0}}{H_0^2}\right| \sim 1 \ .
\end{equation}
Note that Eq. (\ref{e11}) is a \textit{necessary} condition to quit the acceleration and reach to the graceful exit since depending on the functional form of $H_0$, $|\epsilon_1|$ could return to a much less value than unity again. 
Thus, for ending up with the graceful exit, it is essential to restrict $H_0$ so as not to inflate the universe again.

Then, we turn to consider the quantum fluctuation. In the $\phi$-dependent case, the coefficient functions $\fs$ and $\gs$ are given as
\begin{align}
\label{}
    \fs&\simeq {\fs}_0 + {\fs}_1(\widetilde{P_{,\phi}})_0 , \nonumber    \\
    \gs&\simeq {\gs}_0 + {\gs}_1(\widetilde{P_{,\phi}})_0 .
\end{align}
We derive the slow-roll parameter $f_{s1}$ up to $O(\xi)$ as 
\begin{align}
\label{fs1}
    f_{s1}&\simeq \frac{f\left( {\fs}_0' + 9{\fs}_1H_0^2\epsilon_\phi\eta_\phi\right)}{H\left({\fs}_0 + {\fs}_1(\widetilde{P_{,\phi}})_0\right)}  \nonumber \\
    &\simeq\begin{cases}
      \frac{f{\fs}'_0}{H_0{\fs}_0} \qquad \text{for}\quad {\fs}_0 \gg {\fs}_1(\widetilde{P_{,\phi}})_0, \\
      3\eta_\phi\qquad \quad\text{for} \quad {\fs}_0 = 0.
\end{cases}   
\end{align}
We do not need to consider any other cases as long as we impose the ansatz (\ref{ansatz}).
The prime on ${\fs}_0$ denotes the total derivative with respect to $\phi_0$, which is written up to $O(\xi)$ as
\begin{align}
\label{ }
{\fs}_0' :&= \frac{d{\fs}_0}{d\phi_0} \nonumber \\
           &= {{\fs}_0}_{,\phi_0} + {{\fs}_0}_{,\dot{\phi}_0}f_{,\phi_0} + {{\fs}_0}_{,\ddot{\phi}_0}ff_{,\phi_0\phi_0} \nonumber \\
            &+ {{\fs}_0}_{,H_0}{H_0}_{,\phi_0} + {{\fs}_0}_{,\dot{H}_0}f{H_0}_{,\phi_0\phi_0} \ .
\end{align}
$g_{s1}$ is derived in the same way as $f_{s1}$ with replacing ${\fs}_0$ and ${\fs}_1$ to ${\gs}_0$ and ${\gs}_1$.
Thus, if we associate with the ansatz (\ref{ansatz}) and the condition (\ref{eta}), $f_{s1}$ and $g_{s1}$ remain at smaller values than unity. 
Then, to create the spectral tilt consistent to the observations, we must keep at least one of the slow-roll parameters in Eq. (\ref{tau}) while we get rid of the integral consisting of the second order products of slow-roll parameters. We do not specify what conditions we need for realizing this situation since it highly depends on the models. In general, however, the conditions for getting rid of the second order products of slow-roll parameters reduce to constraints for the third or higher order derivatives with respect to $\phi$, e.g. $P_{,\phi\phi\phi}$, while Eq. (\ref{eta})-(\ref{e1}) constrain up to the second order derivatives with respect to $\phi$. Thus, we must impose those constraints independently.

In addition to the above conditions, we must impose the stability condition (\ref{stability}) and a few observational constraints to the spectral index $n_s$, tensor-to-scalar ratio $r$, non-gaussianity parameter $f_{NL}$ etc. to obtain viable models of KDI. It might seem that models of KDI are bounded so tightly that there remain only a few regions for successful models. Practically, however, we can satisfy all of the constraints by tuning a few functions of $\phi$ in the action.
In the case of action (\ref{exa}), for instance, all the above constraints reduce to constraints for the function $K(\phi)$ since, in the action, only $K(\phi)$ depends on $\phi$. If we introduce more functions of $\phi$ into the action, easier it would become to satisfy the constraints. 


\section{Summary}\label{6}

We have constructed the model-independent analytical method for kinetically driven inflation (KDI).
In the analysis, we have evaluated the cosmological background splitting into the inflationary attractor and the linear perturbations around it. Establishing the ansatz that all of the derivatives with respect to the scalar field are so small as to be treated as the perturbations, we have derived the generic equation of motion for the linear perturbation of scalar field.
We have also considered the quantum fluctuation around the background, which has a usual scaling. Given the generic action of fluctuation, we have classified the scalar mode function to three cases by the rank of Hankel function under the approximation discarding the integral of the second order products of slow-roll parameters. 

Using these formalisms, we have investigated the behaviors of background and quantum fluctuation dividing into two cases. First, we have considered the case that the system has the scalar shift symmetry.
In this case, we have obtained a description of the background evolution that the perturbation of scalar field behaves almost as ultra slow-roll inflation around the exact de Sitter attractor. However, the ultra slow-roll motion of perturbation causes rapid changes of the slow-roll parameters, so that we must dismiss all or a few slow-roll parameters from the spectral index to apply our formalisms. As a result, it turns out that all the shift symmetric models of KDI, which are incorporated in our generic framework, cannot create the scalar spectral tilt consistent to the observations. Therefore, to obtain viable models of KDI within our framework, we need to add another source of the scalar fluctuation, or, break the shift symmetry. 
Then, we have considered the shift breaking case. In this case, the background evolution is described as the perturbation of scalar field ``slow-rolls" around the quasi-de Sitter attractor.
Imposing a few theoretical requirements, we have derived the four essential conditions for the shift breaking models of KDI associated with the graceful exit. These conditions generalize the corresponding conditions of the previous models. 


We have obtained the model-independent properties of KDI only with a few assumptions. If new models of KDI based on new scalar-tensor theories are constructed in the future, we can easily discriminate which models are viable or not by using our formalisms as far as the new theories belong in our generic framework.

\section*{Acknowledgement}

This work was supported partially by the National Natural Science Foundation of China under Grant Nos. 11475065 and 11690021.

\appendix

\section{Derivation of the general form of background equations}\label{A}

We start from the action $S[\phi,g_{\mu\nu}]$. Taking the flat FRW background, we obtain the action for the background as
\begin{align}
\label{}
    S_{\rm bg}= \int dtd^3xNa^3P\left[\phi,\,\dot{\phi},\, \ddot{\phi}, \, H,\, \dot{H},\,N,\, \cdots\right]  \ ,
\end{align}
where the ellipsis denotes the higher order time derivative terms.
Differentiating this with respect to $N$, we obtain a Euler-Lagrange equation. That equation must be a constraint equation corresponding to the Fridmann equation, otherwise we have extra degrees of freedom since we cannot determine the evolution of the background with a initial condition for ($\phi,\,\dot{\phi})$ alone. Then, taking the flat FRW solution $N=1$, we obtain the general form of constraint equation as Eq. (1).

Next, we consider the variation of action with respect to $\phi$.
The functional $P$ varies with respect to $\phi$ as
\begin{align}
\label{}
    \delta_\phi P= \frac{\partial P}{\partial \phi}\delta \phi + \sum_{n=1}\frac{\partial P}{\partial \phi^{(n)}}\left(\frac{d}{dt}\right)^n\delta \phi   \ ,
\end{align}
where $\phi^{(n)}$ implies the n-th order time derivative of $\phi$.
The variation of action with respect to $\phi$ can be reduced after a integration by parts to 
\begin{align}
\label{}
   \delta_\phi S_{\rm bg}= &\int dt d^3x\left\{ Na^3\frac{\partial P}{\partial \phi}  \right.\nonumber \\
    &+ \left. \sum_{n=1}(-1)^n\left(\frac{d}{dt}\right)^n\left(Na^3\frac{\partial P}{\partial \phi^{(n)}}\right)\right\}\delta \phi \ .
\end{align}
Differentiating the action with respect to $\phi$ and taking $N=1$, we obtain
\begin{align}
\label{}
    &\frac{d}{dt}\left( a^3J\right)= P_{,\phi} \ ,  \\
    &J:= \sum_{n=1}\frac{(-1)^{n-1}}{a^3}\left(\frac{d}{dt}\right)^{n-1}\left(a^3\frac{\partial P}{\partial \phi^{(n)}}\right)\ .
\end{align}
Thus, we can always get the form of Eq. (2) if we have the flat FRW background solution.

\section{Derivation of the massless action for the scalar fluctuation}\label{B}

In this appendix, we show how to get the massless action (\ref{s2}) for the single field inflation. 
We consider only the case we can use the scalar field $\phi$ as a clock field and take the unitary gauge where 
\begin{align}
\label{}
    \phi = \phi(t) + \delta \phi,\quad \delta \phi = 0 \ . 
\end{align}
We can always take this gauge within the finite time if on the inflationary background, the theory has 2 tensor modes and 1 scalar mode only in the focused energy range. Using the ADM variables \cite{Arnowitt:1962hi} 
\begin{equation}
\label{ }
ds^2 = -N^2dt^2 + h_{ij}(N^idt + dx^i)(N^jdt + dx^j)  \ ,
\end{equation}
we can reduce the original action $S[\phi, g_{\mu\nu}]$ to the unitary gauge action
\begin{align}
\label{sug}
    S_{\rm u.g}= \int dtd^3x \sqrt{h}\,{\cal L}_{\rm u.g}[N,\,K_{ij},\,{}^3R_{ij},\,h^{ij},\,t,\,\cdots] \ .   
\end{align}
$K_{ij}$ and ${}^3R_{ij}$ are the extrinsic curvature and the intrinsic curvature respectively, and the ellipses imply irrelevant terms involved with higher order derivatives. At this point, the unitary gauge action lost the general diff-invariance but still preserves the spatial diff-invariance. 

Conversely, when we have a clock field, we can recover the general diff-invariance by identifying the timelike unit normal $n_\mu$ to the product of clock field \cite{Cheung:2007st, Gao:2014soa} as 
\begin{align}
\label{trick}
    n_\mu&= -\frac{\partial_\mu\phi}{\sqrt{2X}} \ .
\end{align}
This procedure is regarded as the St\"uckelberg trick. If we re-introduce the scalar field as described above, we can replace all of the ingredients in the unitary gauge action to the products of $\phi$ and $g_{\mu\nu}$, for example,
\begin{align}
\label{}
 h_{ij}\rightarrow g_{\mu\nu} + n_\mu n_\nu = g_{\mu\nu} +\frac{\partial_\mu\phi\partial_\nu\phi}{2X}   \ .
\end{align} 
Using the St\"uckelberg trick, we find that we should treat the first order derivatives of $N$ in the unitary gauge action as irrelevant operators. First, we consider 
the acceleration $a_\mu:= n^\nu\nabla_\nu n_\mu = (0,\, \partial_i{\rm ln}N)$. If we recover the general diff-invariance by the trick, 
the terms consisting of acceleration are converted to the higher order derivative terms, for instance, as  
\begin{align}
\label{}
    a_\mu a^\mu = \frac{\nabla_\mu X\nabla^\mu X}{(2X)^2} + \frac{(\partial^\mu\phi\nabla_\mu X)^2}{(2X)^3} \ .
\end{align}
%
It is well-known that in general, those higher order terms introduce extra degrees of freedom which cause the ghost instability and break the unitarity \cite{Woodard:2006nt}. 
Therefore, if we prohibit any accidental cancellations between the higher order derivative terms originated from the acceleration and the other higher order derivative terms, we should ignore all  the terms involved with the acceleration from the unitary gauge action as irrelevant operators. For the same reason, we should ignore the time derivative terms of $N$ also, such as $N_\bot:= \dot{N}-N^i\partial_iN$. Thus, we regard the unitary gauge action provided in Eq. (\ref{sug}) as the most generic action for the single field inflation, which we can obtain naturally.
We will discuss later about the case where the accidental cancellations does occur. 



To fix the residual gauge degrees of freedom, we take the comoving gauge as follows:
\begin{align}
\label{com}
    h_{ij}&= {\rm e}^{2(\rho + \zeta)}[{\rm e}^\mathbf{\gamma}]_{ij}, \nonumber \\
    {\rm e}^\rho &:= a(t),\quad \partial_i\gamma_{ij} = 0=\gamma_{ii}\ ,
\end{align}
where $\zeta$ is the comoving curvature perturbation and $\gamma_{ij}$ is the tensor fluctuation. 
Hereafter, we ignore $\gamma_{ij}$ and set $[{\rm e}^\mathbf{\gamma}]_{ij}$ to $\delta_{ij}$ since we intend to derive the quadratic action of $\zeta$. 
%
Then, the action for scalar fluctuations reduces to
\begin{align}
\label{sc}
    S&= \int dtd^3x\,{\rm e}^{3(\rho+\zeta)}{\cal L}_c[N,\,K^i_{\ j},\,{}^3R^i_{\ j},\, \delta^i_j,\,t,\,\cdots]  \ ,
\end{align}
%
where we rewrote the ingredients of Lagrangian density so as not to include $h^{ij}$ for later convenience. The extrinsic and intrinsic curvatures in the comoving gauge are given as
\begin{align}
\label{K}
    K^i_{\ j}&= \frac{h^{ik}}{2N}\left(\dot{h}_{kj} - 2D_{(k}N_{j)}\right) \nonumber \\ 
    &=\frac{1}{N}\left[ (H+\dot{\zeta})\delta_{ij} \right.\nonumber \\
    &\quad\left. -{\rm e}^{-2(\rho+\zeta)}\left(\partial_{(i}N_{j)}
       -2\partial_{(i}\zeta N_{j)} + \partial_k\zeta N_k\delta_{ij}\right)\right] \ , \\
    \label{R}
    {}^3R^i_{\ j}&= -{\rm e}^{-2(\rho+\zeta)}\left(\partial_i\partial_j \zeta +\partial^2\zeta\delta_{ij}
    -\partial_i\zeta\partial_j\zeta + (\partial\zeta)^2\delta_{ij}\right) \ .
\end{align} 
We do not distinguish $\delta^i_j$, $\delta^{ij}$ and $\delta_{ij}$, and we raise or lower the indices of spatial derivatives by $\delta_{ij}$. $\partial^2$ denotes $\partial_i\partial_i$. 
We note that in the action (\ref{sc}), there does not exist the quadratic term of $N_i$ without spatial differentiation since the unitary gauge action (\ref{sug}) preserves the spatial diff-invariance.  

Differentiating the action (\ref{sc}) with respect to $N$ and $N_i$, we obtain the Hamiltonian constraint and the momentum constraint respectively as
\begin{align}
\label{conN}
    \frac{\delta S}{\delta N}& = {\rm e}^{3(\rho+\zeta)}\frac{\partial {\cal L}_c}{\partial N} = 0 \ ,   \\
    \label{mom}
    \frac{\delta S}{\delta N_i}& = \sqrt{h}h^{(ik}
    D_j\left( \frac{1}{N}\frac{\partial {\cal L}_c}{\partial K^k_{\ j)}}\right) \nonumber \\
    &= {\rm e}^{(\rho+\zeta)}
    D_j\left( \frac{1}{N}\frac{\partial {\cal L}_c}{\partial K^{(i}_{\,\ j)}}\right) = 0\ .
\end{align}
Based on the structure of indices, we can decompose the momentum constraint (\ref{mom}) as 
\begin{align}
\label{mom2}
   &{\rm e}^{(\rho+\zeta)}D_j\left(F\delta^j_i + F_{(i}^{\,\ j)}\right) = 0 \ , \\
    & F=F[N,\,K^i_{\ j},\,{}^3R^i_{\ j},\, \delta^i_j,\,t] ,\nonumber \\
    &F^{\ j}_i = F^{\ j}_i[N,\,K^i_{\ j},\,{}^3R^i_{\ j},\, \delta^i_j,\,t] \ ,\nonumber 
\end{align}
where $F$ is a scalar and $F^{\ j}_i$ is a tensor on the spatial hypersurface. To obtain the quadratic action, we need to solve the constraints only up to the first order of fields. Expanding the Hamiltonian constraint (\ref{conN}) up to the first order of fields, we obtain 
\begin{align}
\label{ham}
    {\rm e}^{3(\rho+\zeta)}\frac{\partial {\cal L}_c}{\partial N} = {\rm e}^{3\rho}(1+3\zeta)\left.\frac{\partial {\cal L}_c}{\partial N}\right|_{\rm 0} + {\rm e}^{3\rho}\left.\frac{\partial {\cal L}_c}{\partial N}\right|_1 = 0   \ ,
\end{align}
where the subscripts denote the order of fields. The zeroth order part, however, corresponds to the constraint equation of background, so it vanishes. Then, from the remaining term, we obtain a generic form of Hamiltonian constraint as
\begin{equation}
\label{ham2}
c_1N_1+ c_2\partial_iN_i + c_3\dot{\zeta} + c_4{\rm e}^{-2\rho}\partial^2\zeta= 0 \ ,
\end{equation}
where $c_{1\sim4}$ are functions of time, and $N = 1 + N_1$. Note that from Eq. (\ref{K}) and (\ref{R}), each spatial derivative and each $N_i$ must be followed with a factor ${\rm e}^{-\rho}$. We write the factor separately only for the spatial derivative of $\zeta$. 

The momentum constraint (\ref{mom2}), up to the first order of fields, can be reduced to
\begin{equation}
\label{ }
\partial_i F + \partial_jF^{\ j)}_{(i} = 0 \ .
\end{equation}
From this, we obtain a generic form of momentum constraint as
\begin{align}
\label{mom3}
    &\partial_i\left(c_2N_1 + c_5\partial_jN_j + c_6\dot{\zeta} + c_7{\rm e}^{-2\rho}\partial^2\zeta\right) \nonumber  \\
      &+ \partial_j\left(c_8\partial_{(i}N_{j)}\right)= 0\ ,  
\end{align}
where $c_{5\sim8}$ are functions of time. 
%
%
%
Since we have already fixed the gauge sufficiently\footnote{We still have gauge symmetries such as the spatial rotation, $x^i\rightarrow x^i + {\epsilon}^i_jx^j$. These residual gauge symmetries, however, does never affect to solvability of the constraint equations.}  at Eq. (\ref{com}), we can  solve the constraints (\ref{ham2}) and (\ref{mom3}) for $N_1$ and $\partial_iN_i$.
If we take a ``physical" solution which is held apart from the spatial infinity, we get
\begin{align}
\label{solution}
    N_1&=N_1[\dot{\zeta},\,{\rm e}^{-2\rho}\partial^2\zeta] \ ,  \nonumber \\
    \partial_iN_i&=(\partial_iN_i)[\dot{\zeta},\,{\rm e}^{-2\rho}\partial^2\zeta]  \ .
\end{align}
%

On the other hand, expanding the action (\ref{sc}) up to the second order of the fields, we obtain the quadratic action consisting of $N$, $N_i$ and $\zeta$. 
 We do \textit{not} have, however, any terms made of $N\zeta$, $\zeta\partial_iN_i$ and $\partial_i\zeta N_i$ in the quadratic action since the generic constraint equations (\ref{ham2}) and (\ref{mom3}) do not include any terms proportional to $\zeta$ without differentiation. Therefore, associated with Eq. (\ref{K}) and (\ref{R}), we can reduce the ingredients of quadratic action to
\begin{align}
\label{ss2}
    S_2^{(S)}
          &= \int dtd^3x{\rm e}^{3\rho} \nonumber \\
          &\quad\times{\cal L}_2[N_1,\,\partial_iN_j,\,\dot{\zeta},\, {\rm e}^{-2\rho}\partial_i\partial_j\zeta,\, {\rm e}^{-2\rho}(\partial\zeta)^2,\,\delta_{ij},\,t]  \ .
\end{align}
Here, ${\rm e}^{3\rho}{\cal L}_2$ implies the second order part of Lagrangian density. 
Substituting the solution (\ref{solution}), we can find that the classical quadratic action for $\zeta$
becomes massless;
\begin{equation}
\label{ }
S_2^{(S)} = \int dt d^3x {\rm e}^{3\rho}{\cal L}_2[\dot{\zeta},\, {\rm e}^{-\rho}\partial_i\zeta,\,{\rm e}^{-\rho}\partial_i,\,t\,] \ .
\end{equation}
If we consider the usual scaling case where $\omega\propto k$, we can restrict the generic form of quadratic action of $\zeta$ to Eq. (\ref{s2}). 

We may extend the same procedure to the non-linear fluctuations \cite{Tanaka:2012wi}. If we can solve the constraints (\ref{conN}) and (\ref{mom}) perturbatively to arbitrary order of the fields and we do never encounter any strong couplings among the fields, we can conclude that at least barely, the non-linear comoving curvature perturbation remains massless for the system described by Eq. (\ref{sug}).  

Lastly,  while it is an unnatural situation, we discuss about the case where the first order derivatives of Lapse function exist in the unitary gauge action as relevant operators. As we mentioned, when we recover the general diff-invariance by the St\"uckelberg trick, we need the accidental cancellations between the higher order derivative terms originated from the derivatives of Lapse function and the other higher order derivative terms to avoid the ghost instability. Such accidental cancellations indeed occur in the kinetic matrix of a class of scalar-tensor theories, so called DHOST theory \cite{Gleyzes:2013ooa, Gleyzes:2014dya, Gleyzes:2014qga, Gleyzes:2014rba, Langlois:2017mxy}. Although DHOST theory includes the first order derivatives of Lapse function in the unitary gauge action, it is showed that the resulting quadratic action for the scalar mode has the same form as Eq. (19) \cite{Langlois:2017mxy}. We have not identified all classes of DHOST theory yet, so that Eq. (19) could cover the known and unknown classes of DHOST theory potentially.


\section{Ultra slow-roll solution}\label{C}

We consider the shift symmetric system which do \textit{not} have any root solutions for $J=0$ and ${\cal E}=0$. Even for this case, we may obtain a quasi-de Sitter solution if the action includes a positive constant potential.
%
When the evolution of the universe is dominated by the potential, we may obtain the following background evolutions:
\begin{align}
\label{FU}
     &H^2\simeq \frac{\alpha V_0}{M_{\rm Pl}^2} = \text{const.}, \\
     &J(\dot{\phi},\, H,\,\cdots)= c_Ja^{-3},
\end{align}
where $\alpha$ is a constant and $V_0$ is the constant potential. 
Since it is supposed that there is no root solution, the behavior of scalar field differs from that of KDI crucially. Assuming that $J$ is expanded by a positive power series of $\dot{\phi}$ and the dominant term of $J$ at a certain time is given by the $p$-th power of $\dot{\phi}$, we obtain
\begin{align}
\label{Jp}
    J&\sim c_p\dot{\phi}^p \ , \\
    \label{usr}
      \dot{\phi}& \propto a^{-\frac{3}{p}},
\end{align}
where $p>0$ and $c_p$ is a constant.
Since the field velocity decreases with the quasi-de Sitter expansion obtained by (\ref{FU}), 
after several e-folds, $J$ is eventually governed by the lowest power of $\dot{\phi}$ \cite{Hirano:2016gmv}.
%
In this mechanics, unlike KDI, the field velocity always decreases towards $0$ in negative powers of $a$, and the scalar field moves quite slowly on the constant potential. This ``ultra" slow motion of scalar field is encoded into the decreasing time derivative of Hubble parameter, which would be written as
\begin{equation}
\label{ }
\dot{H}\propto \dot{\phi}^{q} \propto a^{-\frac{3q}{p}},
\end{equation}
where $q$ is a positive constant. Thus, we still have a slight deviation from an exact de Sitter expansion involved with the quite small field velocity.
Inflation induced by this mechanism is referred to as ultra slow-roll inflation \cite{Kinney:2005vj, Hirano:2016gmv}.
We note that the ultra slow-roll solution cannot become an attractor solution since from Eq. (\ref{usr}), the second order time derivative of scalar field balances with the first order time derivative of scalar field unlike the KDI attractor.
  
Then, we turn to consider the quantum fluctuation. In the shift symmetric case, all the quantities should be expressed by constants and the scale factor $a$ only. We assume that at a certain time, $\fs$ and $\gs$ are written as
\begin{align}
\label{}
    \fs&\sim f_0a^{-3p_f},\quad \gs \sim g_0a^{-3p_g},\nonumber    \\
    c_s^2&\sim \frac{f_0}{g_0}a^{-3(p_f -p_g)}\ ,
\end{align}
where all the characters are assumed to be positive constants. The slow-roll parameters behave as
\begin{align}
\label{}
    \epsilon_1&\propto a^{-\frac{3q}{p}}, \quad f_{s1}= -3p_f,\quad g_{s1} = -3p_g , \nonumber \\ 
    \epsilon_2& \simeq -\frac{3q}{p},\quad f_{s2}= g_{s2}= 0 \ .  
\end{align}
Similarly to the shift symmetric case of KDI, $\epsilon_1$ decreases towards $0$, while $\epsilon_2$ remains almost as a constant. Thus, to apply the approximation where we discard $\epsilon_1\epsilon_2$, we must dismiss $\epsilon_1$ also from the spectral index. If $p_g\geq1$, the scalar fluctuation grows even on superhorizon scales since the parameter $\nu_s$ (\ref{nus}) becomes zero or less than zero. In the original model of ultra slow-roll inflation \cite{Kinney:2005vj} where the action is given by
\begin{equation}
\label{ }
S = \int d^4x\left[ \frac{M_{\rm Pl}^2}{2}R -X -V_0\right] ,
\end{equation}
the parameters are derived as
\begin{align}
\label{ }
p&= 1,\quad q=2, \quad f_0=g_0, \quad p_f=p_g =2, \nonumber \\
\nu_s &= -\frac{3}{2}.
\end{align}
Using Eq. (\ref{nm}) for this model, we have no spectral tilt as well as the type (i) theories of KDI. The scalar fluctuation for this model, however, grows on superhorizon scales since $\nu_s<0$.

\bibliographystyle{apsrmp}

\end{document}